\documentclass[twocolumn,showpacs,preprintnumbers,amsmath,amssymb]{revtex4}

\usepackage{graphicx}
\usepackage{dcolumn}
\usepackage{bm}

\def\ie{i.e.~}

\def\anp#1#2#3{{Ann. Prob.} {\bf #1}, #2 (#3)}

\def\jcp#1#2#3{{J. Chem.  Phys.} {\bf #1}, #2 (#3)}

\def\jmp#1#2#3{{J. Math. Phys.} {\bf #1}, #2 (#3)}
\def\jpa#1#2#3{{J. Phys. A: Math. Gen.} {\bf #1}, #2 (#3)}

\def\philo#1#2#3{{Phil. Mag.} {\bf #1}, #2 (#3)}
\def\phys#1#2#3{{Physica} {\bf #1}, #2 (#3)}

\def\prs#1#2#3{{Proc. R. Soc.} {\bf #1}, #2 (#3)}

\def\pra#1#2#3{{Phys. Rev. A.} {\bf #1}, #2 (#3)}

\def\prl#1#2#3{{Phys. Rev. Lett.} {\bf #1}, #2 (#3)}

\begin{document}

\title{\bf Random Trimer Tilings}

\author{Anandamohan Ghosh\footnote{ananda@theory.tifr.res.in}
and Deepak Dhar\footnote{ddhar@theory.tifr.res.in} }

\affiliation{ Department of Theoretical Physics, 
Tata Institute of Fundamental Research, \\
Homi Bhabha Road, Mumbai 400 005, India.}

\author{Jesper L. Jacobsen\footnote{jesper.jacobsen@u-psud.fr}}
\affiliation{LPTMS, UMR CNRS 8626, Universit\'e Paris Sud, 91405 Orsay, France}
\affiliation{Service de Physique Th\'eorique, URA CNRS 2306, CEA Saclay, 91191 Gif sur Yvette, France}

\begin{abstract}
  We study tilings of the square lattice by linear trimers. For a cylinder
of circumference $m$, we construct a conserved functional of the base of
the tilings, and use this to block-diagonalize the transfer matrix. The
number of blocks increases exponentially with $m$. The dimension of the
ground-state block is shown to grow as $(3 / 2^{1/3})^m$. We numerically
diagonalize this block for $m \le 27$, obtaining the estimate $S_\infty =
0.158520 \pm 0.000015$ for the entropy per site in the thermodynamic
limit.  We present numerical evidence that the continuum limit of the
model has conformal invariance.  We measure several scaling dimensions,
including those corresponding to defects of dimers and $L$-shaped trimers.  
The trimer tilings of a plane admits a two-dimensional height
representation. Monte Carlo simulations of the height variables show that
the height-height correlations grows logarithmically at large separation,
and the orientation-orientation correlations decay as a power law.
\end{abstract}

\pacs{ }

\maketitle

\section{Introduction}

The study of statistics of densely packed polymers has long been of interest
to physicists. Onsager had argued that solutions of long rod-like molecules
should show orientational order at high densities \cite{onsager}. Flory's
approximate analysis suggested that linear rod-like molecules on a two-dimensional lattice
should also exhibit an orientational order at high densities \cite{Flo56}.
However, for the case of {\em dimers} on a lattice---the only case that is
analytically soluble---it is known that for all non-zero monomer densities,
there is no long-range orientational order \cite{Hei72}. In the limit of zero
monomer density, one gets power-law decay of correlations for (bipartite)
square and hexagonal lattices \cite{Fis61, Kas63}, but only short-ranged
correlations on the triangular lattice \cite{Fen02}. 
Recently there have been studies of the dimer problem on the cubic lattice \cite{Hus03}
and of interacting classical dimers on the square \cite{Ale05} and cubic \cite{Ale06} lattices.

Monte-Carlo simulations
of Baumg\"{a}rtner show that for semi-flexible lattice polymers close to full
packing there is no long range order, no phase transition, and the correlation
length is of the order of the size of the polymer \cite{Bau84}. However, the
exact solution of a {\em single} semi-flexible polymer that is {\em fully}
packed on the square lattice (the so-called Flory model) exhibits a
low-temperature phase of crystalline order and an infinite-order transition to
a disordered, critical high-temperature phase in which the critical exponents vary
continuously with temperature \cite{JK}. 

It is generally believed, but not 
proved, that in the continuum
case in three dimensions, long needle-like molecules would undergo a
isotropic-nematic transition. In two dimensions, a spontaneous breaking of
continuous rotational symmetry is not allowed, but there is a
Kosterlitz-Thouless phase with power-law decay of orientational correlation
functions \cite{Fre85,Kha05}.
But the situation is less clear for systems of hard-core molecules on a
lattice. In the limit of high density, one can get a
solid-like phase where one of the sublattices is preferentially occupied,
e.g., in the cases of hard squares and hard hexagons \cite{baxter}. The
same behavior is seen in Monte Carlo studies with lattice models of
extended hard discs \cite{extendedcore}. However, triangular trimers on
the triangular lattice can be solved exactly, and do not show a long-range
order even at close packing \cite{Ver99}. De Gennes has argued that long
straight needles may not show an ordered phase on the square lattice
\cite{degennes}. There are not many studies of other molecular shapes in
lattice models, as realistic modeling of actual experimental system
assemblies of different shaped molecules (e.g., ellipsoids, banana-shaped
molecules, etc.) is better done in the continuum space.  Tilings by
L-shaped trimers and T-shaped tetramers of $m \times \infty$ strips, for
$m \leq 5$ have been studied using the transfer matrix technique earlier
\cite{CMpo,Fro96}.

\begin{figure}
\includegraphics[scale=.8, angle=0]{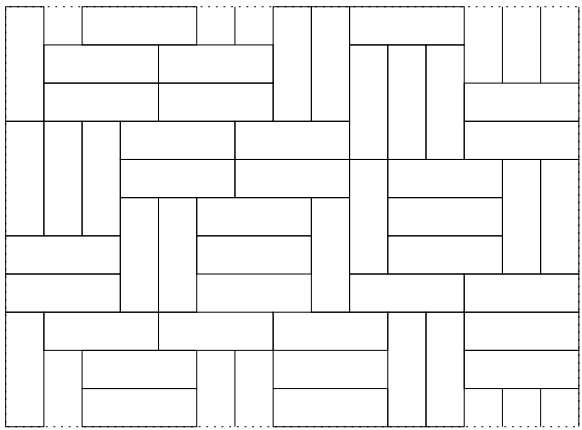} 
\caption{A  trimer tiling with periodic boundary condition in 
vertical direction.} \label{figure1}
\end{figure}

In this paper we study random trimer tiling of the square lattice with
horizontal and vertical trimers (Fig. \ref{figure1}). We wish in particular
to assess if there is a long-ranged correlation of the orientational order
in this problem in the limit of zero monomer density. The problem is first
addressed in the geometry of semi-infinite cylinders of size $m \times
\infty$. We show how to set up the corresponding transfer matrix and
numerically diagonalize it for $m \le 27$. We use these data to
extrapolate to the $m \to \infty$ limit, determining in particular the
entropy per site in a random trimer tiling in the thermodynamic limit. We
also prove the existence of a family of matrices which commute with the
transfer matrix and show that the transfer matrix decomposes into a number
of blocks which is exponentially large in $m$. Our results for the free
energy and various correlation functions are consistent with  a 
conformally
invariant system of central charge $c=2.15 \pm 0.2$. In particular,
the correlation length $\xi_m$ is found to increase linearly with $m$.

It is known that trimer tilings admit a two-component height field
representation \cite{Ken00}. In the plane, we use this to study correlations
of the height field by Monte-Carlo simulations, and find that the correlation
function $\left \langle h({\bf r})h({\bf 0})) \right \rangle$ varies as
$\log(|{\bf r}|)$ for large $ |{\bf r}|$. We have also studied the
orientational correlations in the simulations, and find that they decay as
power laws.

The plan of the paper is as follows: 
in Section~\ref{sec:height} we present the construction of a height representation
for a trimer tiling; 
in Section~\ref{sec:motion} the constants of motion are obtained for trimer tilings 
on a cylinder;
in Section~\ref{sec:sectors} the number of disjoint sectors is calculated by the
generating function formalism;
in Section~\ref{sec:transfer} we set up the transfer matrix for the problem,
and in Section~\ref{sec:diag} we present the results of numerical diagonalization
of the transfer matrix;
in Section~\ref{sec:corr} we discuss the trimer-trimer correlation functions 
and in Section~\ref{sec:mc} height-height correlation functions,
both obtained by Monte-Carlo simulations. Our conclusions are
presented in Section~\ref{sec:conclusion}. An appendix adapts the working
of Section~\ref{sec:sectors} to a more general tiling problem.

\section{Trimer tilings and height representation}
\label{sec:height}

\begin{figure}
\includegraphics[scale=.45, angle=0]{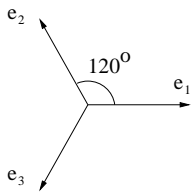} 
\caption{ Choice of unit vectors } 
\label{figure2}
\end{figure}

We represent each trimer as a set of three consecutive {\it squares} along
a line, oriented horizontally or vertically.  For each tiling of a
two-dimensional plane by trimers (Fig. 1), one can define a configuration
of a height model at the {\it vertices} of the square lattice, where the 
heights are two-dimensional vectors
\cite{Ken92,CMht} as follows:

Choose two-dimensional vectors ${\bf e}_1$, ${\bf e}_2$ and ${\bf e}_3$ such
that they satisfy the condition ${\bf e}_1 + {\bf e}_2 + {\bf e}_3 = 0$. A
convenient choice is ${\bf e}_1 = (1,0), {\bf e}_2 = \frac{1}{2}( -1,
\sqrt{3}), {\bf e}_3 = \frac{1}{2}( -1, -\sqrt{3})$, as shown in
Fig.~\ref{figure2}. Equivalently the vectors can be represented as complex
numbers ${\bf e}_1 = 1$, ${\bf e}_2 = \omega$ and ${\bf e}_3 = \omega^2$,
where $\omega={\rm e}^{2\pi i/3}$. The height  ${\bf h}(i,j)$ at any
site $(i,j)$ is an integer linear combination of basis vectors ${\bf 
e}_1$, ${\bf e}_2$ and ${\bf e}_3$. 

\begin{figure} \includegraphics[scale=.9,
angle=0]{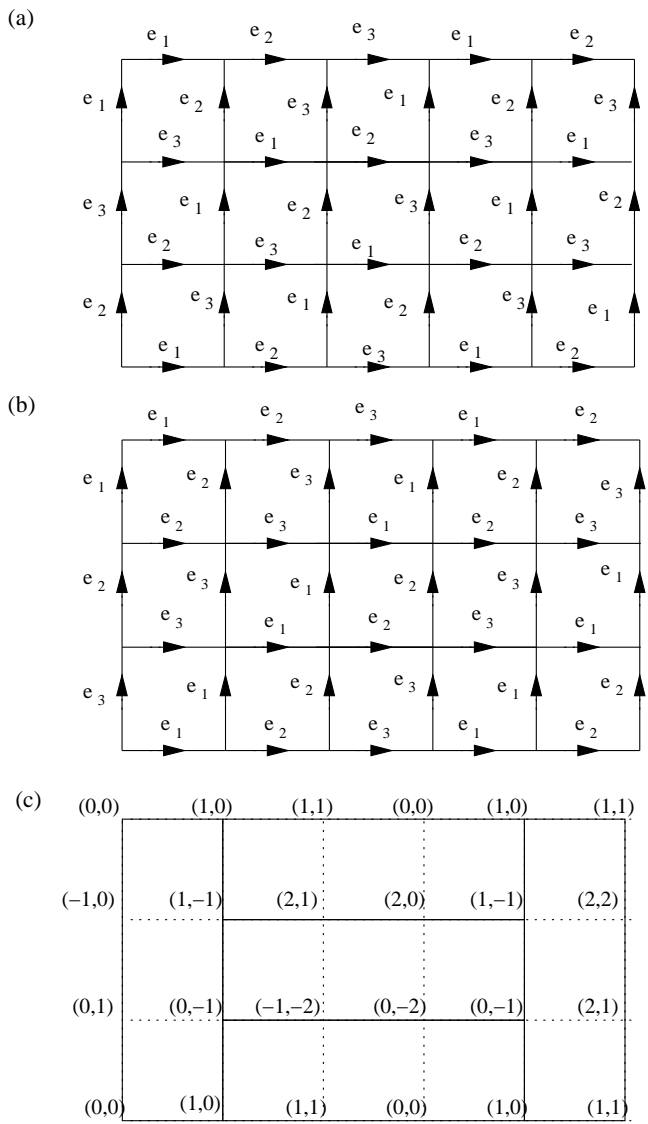} \caption{(a) A trimer tiling of $3 \times 5$ lattice
with assignment of labels to the edges of the lattice (b) An alternative
assignment of labels (c) the height configuration corresponding to label
in (b), where $ (n_1, n_2)$ denotes the height ${\bf h}(n_1,n_2) \equiv
n_1 {\bf e}_1 + n_2 {\bf e}_2$. } \label{figure3} \end{figure}

  The lattice edges
are assumed to be oriented rightwards or upwards, and they 
are labeled
with vectors ${\bf e}_1$, ${\bf e}_2$ and ${\bf e}_3$ periodically as
shown in Fig.~\ref{figure3}(a) and (b). The labeling is such that moving
along a horizontal or vertical line from any vertex up or right encounters
a periodic sequence of labels ${\bf e}_1$, ${\bf e}_2$, ${\bf e}_3$. This
rule still leaves some freedom in choosing the sequence of bonds; two
convenient choices are shown in Figs.~\ref{figure3}(a) and (b).

Now, for any given tiling of the square lattice, the height field ${\bf
h}(i,j)$ is
defined so that if the directed edge $\ell$ from site ${\bf a}$ to the site
${\bf b}$ does not belong to the interior of a tile (i.e., it forms part of
the boundary shared by two tiles), and it has label ${\bf e}_{\alpha}$, then
$h({\bf b}) -h({\bf a}) ={\bf e}_{\alpha}$. This determines the heights at all
vertices up to an unimportant additive constant. The constant can be fixed by
arbitrarily choosing $h({\bf 0}) = {\bf 0}$. 

Note that for the two choices of edge-weights shown in Fig.~\ref{figure3},
the height difference along any edge of trimer has modulus $1$. For (a),
it value is $\sqrt{7}$ or $2$ for the internal edges of trimers, and for
the choice (b), it only takes the value $\sqrt{7}$.  An example of the
values of the height field following the convention (a)  is illustrated in
Fig.~\ref{figure3}(c) for a particular configuration of trimers.
 
\begin{figure}
\includegraphics[scale=1, angle=0]{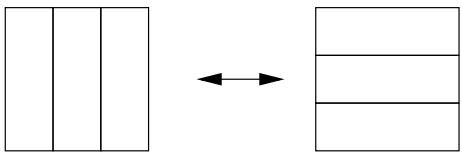} 
\caption{The flip move exchanging three vertical trimers with three 
horizontal trimers.
} \label{figure4}
\end{figure}

\begin{figure}
\includegraphics[scale=.52, angle=0]{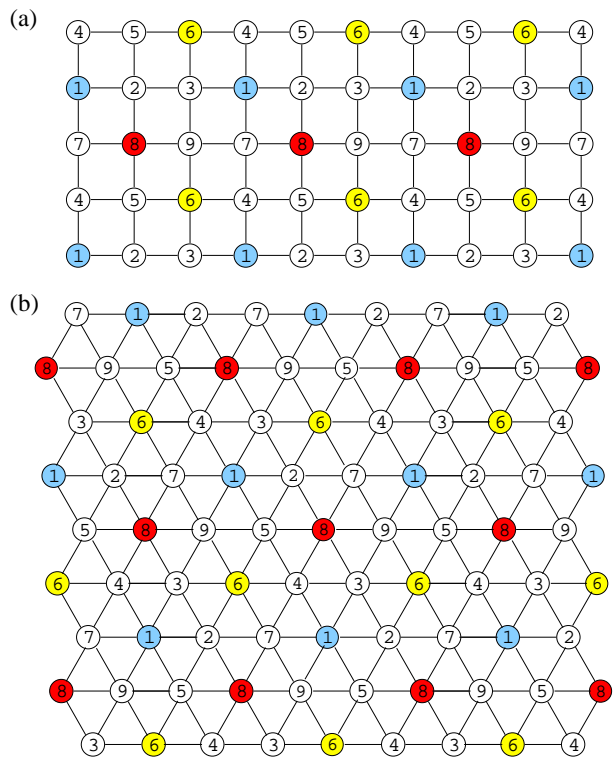}
\caption{(a) $3 \times 3$ superlattice decomposition of the square lattice
into 9 sublattices. (b) Height variables corresponding to a tiling of the
square lattice form a triangular lattice. }
\label{figure5}
\end{figure}

 The choice (a) has a particular advantage. It can be shown that any
trimer covering of a plane can be obtained from any other by a sequence of
the basic flip operation, in which three adjacent horizontal trimers are
replaced by three adjacent vertical ones (Fig. \ref{figure4}). Under a basic
flip move, it is easy to check that the height changes only at four sites and
the modulus of the change in height  $\| \Delta({\bf h}) \|$ is always $
3$. We can think of the values of the height field as forming a triangular
lattice on the complex plane. This lattice can be broken into $9$
sublattices (Fig.  \ref{figure5}), such that even after any such flip, the
value of height stays on the same sublattice. Also, different sublattices
of the height field are in one-to-one correspondence with a $9$ sublattice
decomposition of the original square lattice (Fig. \ref{figure5}).

\section{Constants of motion for trimer tilings on a cylinder}
\label{sec:motion}

In this section we consider the geometry where the square lattice has been
wrapped on a cylinder of circumference $m$. The fully-packing constraint on
the trimer tilings then implies strong constraints on the different
configurations of trimers along a row of length $m$. In fact, given some local
configuration along a row, many local configurations along a different row are
disallowed. If we think of the row-to-row transfer matrix as an trimer evolution
operator for configurations on a line these constraints can be described in terms of some constants of
motion under this evolution.

The simplest constant of motion of this type can be constructed in terms of
the invariants for loops \cite{thurston}. Given a trimer tiling, define an
allowed loop as a sequence of nearest neighbor bonds on the lattice that
returns to the starting point, and does not intersect itself, and none of the
steps crosses a trimer. In other words, an allowed loop goes along the
boundaries of trimers. Define ${\cal G}$ as the group generated by two
generators ${\bf a}$ and ${\bf b}$, which satisfy
\begin{eqnarray}
{\bf a}^3 {\bf b} = {\bf b} {\bf a}^3 \nonumber\\ 
{\bf a} {\bf b}^3  = {\bf b}^3 {\bf a}. \label{groupG}
\end{eqnarray}
Now we attach weights (elements from the group ${\cal G}$) to each step of the
loop. These read respectively ${\bf a}$, ${\bf a}^{-1}$, ${\bf b}$, and ${\bf
b}^{-1}$ for a step of the loop to the right, to the left, up, and down. Finally,
to each loop $L$ we assign a weight $w_L \in {\cal G}$ as the ordered product
of the weights attached to the steps along the loop. Eqs.~(\ref{groupG}) imply
that $w_L$ does not depend on the way (starting point and direction) in which the
loop $L$ is traversed when building up the product weight. Moreover, it is
easy to see that for a non-winding loop (i.e., an allowed loop on the cylinder
which is homotopic to a point), $w_L$ is equal to identity. (The proof goes by
induction on the number of trimers enclosed by the loop.)

Also, $w_L$ takes the same value for any allowed loop that winds around the
cylinder. Indeed, let $L$ be such a loop. Then it is easy to see that $w_L$
does not change if the loop is deformed locally so that the number of trimers
below it changes by $\pm 1$. Thus, $w_L$ is a constant for the tiling.
However, given two products $w_L$ and $w_{L'}$ of the generators ${\bf a}^{\pm
1}$, ${\bf b}^{\pm 1}$, checking whether $w_L = w_{L'}$ by using the rules
(\ref{groupG}) is nontrivial. We now describe a different construction that is
equivalent to this, but more convenient to use.

\begin{figure}
\includegraphics[scale=.45, angle=0]{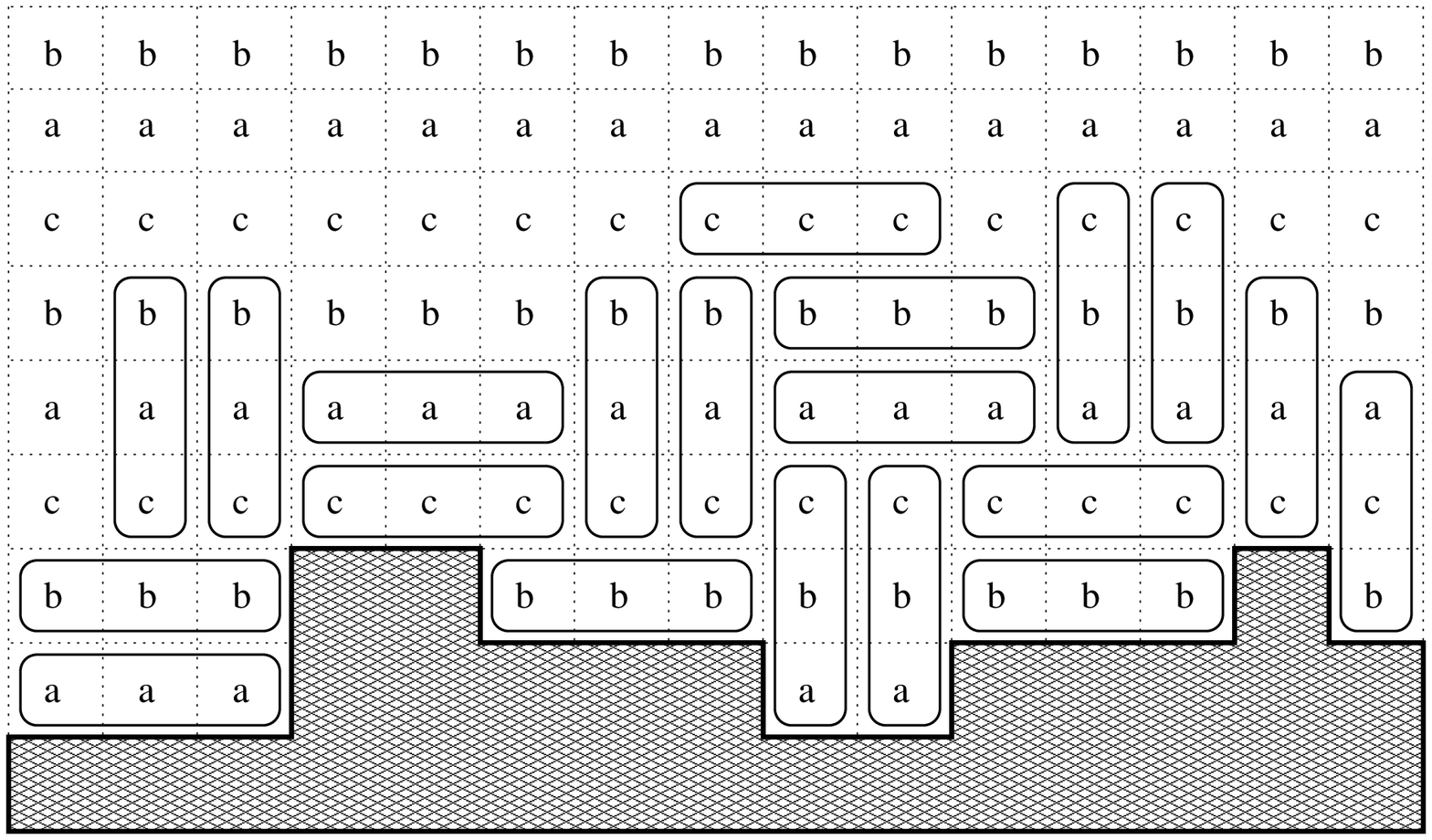} 
\caption{Partial filling on top of a given base (shown hatched).}
\label{figure6}
\end{figure}

Fig.~\ref{figure6} shows a partial tiling of the plane, starting from a base
(shown hatched in the figure). Each trimer occupies three 
horizontally or vertically consecutive squares. 
Periodic boundary conditions in the
horizontal ($x$) direction are assumed. The base is supposed without
overhangs, and can therefore be specified by its height profile, $H_0(x)$. The
only constraints about the partial tiling we assume is that it has no holes,
and no overhangs. In particular, the tiled region can also be described by a
height profile $H(x)$.

We now construct an invariant of the tiling as follows. First, let ${\bf A}$,
${\bf B}$ and ${\bf C}$ be any three non-commuting matrices satisfying the
condition
\begin{eqnarray} \label{eq1}
{\bf A}^3 = {\bf B}^3 = {\bf C}^3 \,.
\end{eqnarray}
Clearly such matrices exist; one family of possible choices is given by
\begin{eqnarray}
{\bf A} &=&  
\left[ 
\begin{array}{ccc}
1 & \lambda_1 & \mu_1 \\
0 & \omega & \nu_1 \\
0 & 0 & \omega^2 
\end{array}
\right] \nonumber \\
{\bf B} &=& 
\left[ 
\begin{array}{ccc}
1 & \lambda_2 & \mu_2 \\
0 & \omega^2 & \nu_2 \\
0 & 0 & \omega 
\end{array}
\right]  \nonumber\\ 
{\bf C} &=&
\left[ 
\begin{array}{ccc}
1 & 0 & 0 \\
\lambda_3 & \omega & 0 \\
\mu_3 & \nu_3 & \omega^2 
\end{array}
\right] \label{specificchoice}
\end{eqnarray}
where $\omega={\rm e}^{2 \pi i/3}$, and $\lambda_i$, $\mu_i$, $\nu_i$ are any
complex numbers.

Next, we assign to each square one of the three colors $a$, $b$ or $c$, where
each color corresponds to a value of its vertical coordinate $y \mbox{ (mod 3)}$ (see
Fig. \ref{figure6}). The height profile $H(x)$ can then be characterized by a
word $l_1 l_2 \cdots l_m$ over the letters $a$, $b$, $c$, where for any
$x=1,2,\ldots,m$ the letter $l_x$ specifies the color corresponding to
$y=H(x)$. For example, the word characterizing the partial tiling in
Fig.~\ref{figure6} reads $bbbaaabcccbccba$. Correspondingly to this word, we
construct a functional of $H(x)$, denoted by ${\mathcal J}(H(x))$, which is
defined as the trace of a product of matrices ${\bf A}$, ${\bf B}$, ${\bf C}$,
where each factor is obtained from a letter in the word by replacing $a
\mapsto {\bf A}$, $b \mapsto {\bf B}$, $c \mapsto {\bf C}$. For example, for
the sequence given above, we get the functional
\begin{equation}
{\mathcal J}(H(x)) =  {\rm Tr} \, [{\bf B B B A A A B C C C B C C B A}] \,.
\end{equation}
Note that by the usual cyclic properties of the trace, the functional depends
only on the height profile, and not on the starting point. We can
therefore write ${\mathcal J}(H(x)) = {\mathcal J}(H)$.

We now have the remarkable theorem: 
\begin{equation}
 {\mathcal J}(H) = {\mathcal J}(H_0) \,.
\end{equation}
In other words, {\em ${\mathcal J}$ is the same for all valid partial tilings,
grown from the same base, and is equal to its value for the base.}

The proof is by induction on the number of trimers in the tiling. It is
clearly true for no trimers. If we add a vertical trimer at $x$, the height
increases by $3$, and the letter $l_x$ does not change. If we add a horizontal
trimer, it must be done on a locally flat substrate. Then in the word, the
substring $aaa$ may be replaced by $bbb$, or $bbb$ by $ccc$, or $ccc$ by
$aaa$. But by Eq.~(\ref{eq1}), this does not change ${\mathcal J}$. {\bf
Q.E.D.}

For the particular choice of ${\bf A}$, ${\bf B}$, ${\bf C}$ given in
Eqs.~(\ref{specificchoice}), we actually have the additional property ${\bf
A}^3 = {\bf B}^3 ={\bf C}^3 = {\bf I}$. For any choice of matrices having this
additional property the word $l_1 l_2 \cdots l_m$ can be transformed into an
``irreducible word'' \cite{DMH} by recursively deleting any subsequent (modulo
$m$) occurrences of three equal letters [i.e., $l_i l_{i+1} l_{i+2} = aaa$, or
$bbb$, or $ccc$] in the original word. In the example shown in
Fig.~\ref{figure6}, the base is characterized by the (reducible) word
$cccbbaaaccaaaba$, which then corresponds to the irreducible word $bbccba$. By
the theorem, this is also the irreducible word of any partial tiling built on
that base.

The invariant ${\mathcal J}$ is a polynomial function of the parameters $
\lambda_i, \mu_i, \nu_i$ in the matrices (\ref{specificchoice}). We can
therefore expand ${\mathcal J}$ in multivariable power series in these
variables. By the theorem, all the coefficients in this expansion are the same
for any allowed tiling, and are constants of motion for the transfer matrix.

For trimers tilings on a torus, we can define other invariants ${\mathcal
J}_{(n_x,n_y)}$ corresponding to other homotopy classes of non-contractible
loops. The allowed homotopy classes are characterized by the winding numbers
$n_x$, $n_y$ along the two coordinate directions, with $n_x \wedge n_y = 1$
and $n_x + n_y \ge 1$ \cite{FSZ87}. Our previous invariant is ${\mathcal
J}={\mathcal J}_{(1,0)}$. However, the invariants ${\mathcal J}_{(n_x,n_y)}$
are not all independent, as can easily be inferred from the simple example
when ${\mathcal J}_{(1,0)}$ corresponds to $abababab\ldots$; and 
there is then only
one possible trimer tiling which completely fixes the values of all other loop
variables.

\section{Decomposition of phase space into disjoint sectors}
\label{sec:sectors}

The word associated with the base can be any sequence of $m$ letters chosen
from $a$, $b$, $c$. The number of all possible base profiles $H_0(x)$ (modulo
the addition of vertical trimers) for a cylinder of width $m$ is therefore
$3^m$.
 
We have already seen that two height profiles $H_1(x)$ and $H_2(x)$ are
reachable from each other if and only if ${\mathcal J}(H_1(x)) = {\mathcal
J}(H_2(x))$. Hence the transfer matrix for this problem has a block diagonal
structure, with no transition possible between configurations with different
${\mathcal J}$.

It is straightforward to determine the number of irreducible words of length
$n \ge 2$ that start with any two specified letters. Let us denote by $N_d(n)$
(resp.\ $N_s(n)$) the number of irreducible words in which the initial two
letters take fixed {\em different} (resp.\ {\em same}) values. Using the
condition that in an irreducible word we cannot have any three consecutive
letters being identical, it is easy to see that they satisfy the recursion
relation
\begin{eqnarray}
N_{d}(n+1) &=& 2 N_d(n) + N_s(n) \nonumber\\
N_{s}(n+1) &=& 2 N_d(n) \label{recursrels}
\end{eqnarray}
for $n \ge 2$, with the initial condition $N_d(2) = N_s(2) = 1$. 
These equations are easily solved giving
\begin{eqnarray}
N_d(n) &=& \frac{1}{4 \sqrt{3} }
[ - (1 - \sqrt{3})^n + (1 + \sqrt{3})^n] \label{solrecrel}  \\
N_s(n) &=& \frac{1}{12} [(3 + \sqrt{3}) (1 - \sqrt{3})^n 
+ (3 - \sqrt{3}) (1 + \sqrt{3})^n]  \nonumber
\end{eqnarray}
The total number of irreducible strings $t_n$
of length $n$ is then obtained by taking into account the multiplicities due
to the possible choices of the two initial letters:
\begin{equation}
 t_n = 6 N_d(n) + 3 N_s(n) \,. \label{multipli}
\end{equation}

The total number of sectors for a cylinder of width $m$ (with $3 | m$) is then
$\sum_{k=0}^{m/3} t_{3k}$. Note that this number is exponentially large in $m$.
For comparison, for dimer tilings (with $2 | m$) the number of sectors is just $2m+1$.

Note that each of the $t_m$ sectors corresponding
to an irreducible word of length $m$ contains only one state of the transfer
matrix (these sectors are ``stuck''). On the other hand, the $t_0=1$ sector
corresponding to the empty irreducible word comprises a number of states in
the transfer matrix which grows exponentially with $m$.

It is straightforward 
to obtain precisely this latter number, i.e., the number of
words of $n$ letters that reduce to the empty irreducible word. Namely, it
corresponds to the dimension of the transfer matrix in the ground state
sector, i.e., the sector from which the free energy is obtained.

To this end, let us first define
\begin{eqnarray}
G_a &=& aaa + abbbaa + aabbba + acccaa \nonumber\\
    &+& aaccca + abbbcccaa + acccbbbaa + ...    \label{gaaa}
\end{eqnarray}
as the formal sum over all unfactorizable words with initial letter $a$ that
are reducible to the empty word. Here unfactorizable means that the words
contributing to $G_a$ must not be the concatenation of two non-empty words
each of which is in turn reducible to the empty word. Note also that the
requirement that the initial letter be $a$ implies, by the property of
unfactorizability, that the last reduction before reaching the empty word is
of the type $aaa \mapsto \emptyset$. 
We similarly define $G_b$ and $G_c$. The sum in
Eq.~(\ref{gaaa}) can then be expressed as
\begin{eqnarray}
G_a &=& a [1+(G_b+G_c)+(G_b+G_c)^2+ \cdots]        \nonumber\\
    &~& a [1+(G_b+G_c)+(G_b+G_c)^2+ \cdots] a      \nonumber\\
&=& a \frac{1}{1 - (G_b + G_c)} a \frac{1}{1 - (G_b + G_c)} a \,.
\label{Ga}
\end{eqnarray}
Now substituting $a = b = c = x$ in Eq.~(\ref{gaaa}) we obtain the
generating function for irreducible words with a formal weight $x$ per letter:
\begin{eqnarray}
g(x) = G_a(a=x) = \sum_{n=1}^{\infty} g_{3 n} x^{3n} \,,
\end{eqnarray}
where $g_{3n}$ is the number of different unfactorizable words that start with
a given fixed letter and are reducible to the empty word. By Eq.~(\ref{Ga}),
$g(x)$ then satisfies the equation
\begin{eqnarray}
g(x) [1 - 2 g(x)]^2 = x^3 \,.  \label{galp}
\end{eqnarray}
This is a cubic equation in $g(x)$ and can be solved explicitly. Among the
three solutions for $g(x)$, two can be discarded as unphysical on the ground
that $g(0) \neq 0$. The last, physical solution can be expanded into a
polynomial series in $x$, as
\begin{eqnarray}
g(x) &=& x^3 + 4 x^6 + 28 x^9 + 240 x^{12} + 2288 x^{15} \nonumber\\
  &+& 23296 x^{18} + 248064 x^{21} + 2728704 x^{24} + \cdots \,. \nonumber
\end{eqnarray} 
Apart from the trivial root in $x=0$, $g(x)$ has two non-trivial coincident
roots for $x^3 = x_{\rm c}^3 = 2/27$. For $x$ near $x_{\rm c}$, $g(x)$ varies
as
\begin{eqnarray}
g(x=x_{\rm c} - \delta) = \frac{1}{6} - A \delta^{1/2} + o(\delta^{1/2}) \,.
\end{eqnarray}
This implies that for large $n$
\begin{eqnarray}
g_{3n} \sim A \left( \frac{27}{2} \right)^n \frac{1}{n^{3/2}}.
\label{gnlarge}
\end{eqnarray}

Using $g(x)$, we can now construct the generating function $H(x)$ of all words
(factorizable or not, and with any initial letter) that are reducible to the
empty word. We have clearly
\begin{eqnarray}
H &=& 1 + (G_a + G_b + G_c) + (G_a + G_b + G_c)^2 + \cdots \nonumber\\
  &=& \frac{1}{1 - (G_a + G_b + G_c)}       \label{gfun}
\end{eqnarray}
Setting $a=b=c=x$ in Eq.~(\ref{gfun}) as before, we get
\begin{eqnarray}
H(x) = \sum_{n=1}^{\infty} H_{3 n} x^{3n} \,,
\end{eqnarray}
where $H_{3 n}$ is the total number of words of length $n$ that are reducible
to the empty word. The leading terms of the polynomial series are
\begin{eqnarray}
H(x) &=&  \frac{1}{ 1 - 3 g(x)} \label{Hcoefs} \\
 &=& 1 + 3 x^3 + 21 x^6 + 183 x^9 + 1773 x^{12} + 18303 x^{15} \nonumber\\
  &+& 197157 x^{18} + 2189799 x^{21} + 24891741 x^{24} + \cdots \,. \nonumber
\end{eqnarray}
For $x$ near $x_{\rm c}$, $H(x) \approx H(x_{\rm c}) - A (x_{\rm c} -
x)^{1/2}$, whence $H_{3 n}$ also varies as
\begin{eqnarray}
H_{3n} \sim A \left( \frac{27}{2} \right)^n \frac{1}{n^{3/2}}.
\end{eqnarray}
for large $n$, where $A$ is the same constant as in Eq.~(\ref{gnlarge}).

The coefficients appearing in Eq.~(\ref{Hcoefs}) coincide with the observed
dimension of the transfer matrix in the ground-state sector (see
Table~\ref{tab1} below).

Finally, let us note that the working of this section can be adapted to a more
general tiling problem. This is relegated to Appendix A.

\section{Setting up the transfer matrix}
\label{sec:transfer}

The number of ways $N_n(m)$ to tile a cylinder of width $m$ (with $3|m$) and
height $n$, with periodic boundary conditions in the $m$-direction and free
boundary conditions in the $n$-direction, can be found as $\langle 0 | (T_m)^n
| 0 \rangle$. Here, $T_m$ is the transfer matrix that adds one row of width
$m$, $|0\rangle$ is an initial state corresponding to an initial horizontal
base, and $\langle 0|$ is a projection operator on the state corresponding to
a final horizontal base. For finite $n$, diagonalizing $T$ by means of a
similarity transformation leads to an expression of the form $N_n(m) = \sum_i
\alpha_i(m) [\lambda_i(m)]^n$, where $\lambda_i(m)$ are the eigenvalues of
$T$. The corresponding amplitudes $\alpha_i(m)$ depend on our choice of
boundary conditions in the $n$-direction. We are mainly interested in the
limit $n \gg 1$, for which one has simply
\begin{eqnarray}
N_n(m) \approx [\lambda_1(m)]^n 
\end{eqnarray}
where $\lambda_1(m)$ is the largest eigenvalue of $T_m$.
The corresponding entropy per site is then
\begin{eqnarray}
S_m = \frac{1}{m} \log \lambda_1(m) \,.
\end{eqnarray}

We shall now describe two different ways of constructing the transfer matrix.

\subsection{First construction}
\label{sec:1stTM}

The first construction is conveniently described in terms of a more
general tiling problem (cf.~Appendix A) of tiling  the plane  by
horizontal $p$-mers (of size $p \times 1$ elementary squares) and vertical
$q$-mers (of size $1 \times q$). The trimer case is recovered for $p=q=3$.

We first shift the tiling by half a lattice spacing, both horizontally and
vertically, with respect to the underlying square lattice. The tile boundaries
then intersect some of the lattice edges, and it is natural to describe the
tiling by assigning an appropriate variable (`spin') $s_i$ to each lattice
edge $i$.

An edge $i$ intersecting a tile boundary has $s_i = 0$. Each vertical $q$-mer
encloses $q-1$ lattice edges which are not intersected by its boundaries; the
lowest of these edges has $s_i=1$, the second-lowest $s_i=2$, and so on, and
the highest edge has $s_i=q-1$. Each horizontal $p$-mer encloses $p-1$ lattice
edges which are not intersected by its boundaries; the leftmost of these edges
has $s_i=q$, the second-leftmost has $s_i=q+1$, and so on, and the rightmost
edge has $s_i=q+p-2$.

\begin{figure}
\includegraphics[scale=.45, angle=0]{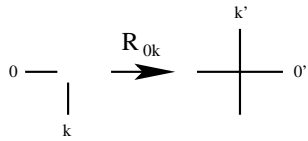} 
\caption{Labeling of the $R$ matrix.} \label{fig:R}
\end{figure}

Using a construction well-known from the theory of integrable systems,
the row-to-row transfer matrix can be written
\begin{equation}
 T_m = {\rm Tr}_0 \, R_{0m} \cdots R_{02} R_{01} \,,
 \label{TRR}
\end{equation}
where each of the matrices $R_{0k}$ act on two ingoing edges (labeled $0$ and
$k$), joins them by adding one vertex, and produces two outgoing edges
(labeled $k'$ and $0'$), as shown in Fig.~\ref{fig:R}. The trace over the
edge $0$ corresponds, graphically, to adding the first horizontal edge of a
new row, and making sure that it joins to a horizontal edge carrying the same
spin $s_0$, once the row has been completed.

It remains to specify the elements of the matrix $R_{0k}(s_0,s_k;s_k',s_0')$.
These are one for the cases
\begin{equation}
 (s_0,s_k;s_k',s_0') = \left \lbrace
 \begin{array}{l}
  (0,0;1,0), (0,0;q,0) \\
  (0,i;i+1,0) \mbox{ with } 1 \le i \le q-2 \\
  (0,q-1;0,0) \\
  (i,0;0,i+1) \mbox{ with } q \le i \le q+p-3 \\
  (q+p-2,0;0,0)
 \end{array}
 \right.
 \label{Rmatrix}
\end{equation}
and zero for all other cases. Indeed, the first line in Eq.~(\ref{Rmatrix})
corresponds to the lower left corner of any tile; the second line to the
interior of a vertical tile; the third line to the upper right corner of a
vertical tile; the fourth line to the interior of a horizontal tile; and the
fifth line to the upper right corner of a horizontal tile. Finally, the trace
in Eq.~(\ref{TRR}) is over $s_0=0,q,q+1,\ldots,q+p-2$.

In some applications it might be of interest to give different weights to horizontal
and vertical tiles. This can readily be done, by attributing the desired weight to the lower
left corner of each tile, corresponding to the first line of Eq.~(\ref{Rmatrix}).

The transfer matrix $T_m$ is constructed in the base of spins states
$(s_1,s_2,\ldots,s_m)$ corresponding to an initial row of vertical edges,
with each $s_k \in \{0,1,\ldots,q-1\}$. The factorization (\ref{TRR}) of $T_m$ is
particularly suited for using sparse-matrix and hashing techniques, so that
$T_m$ can be multiplied onto an arbitrary initial vector in time $\sim m \,
{\rm dim}(T_m)$. The first few eigenvalues of $T$ can be found by an iterative
scheme (the so-called power method \cite{Wilkinson}) based on iterating such
multiplications.

The ground-state sector, corresponding to growing the tiling from a horizontal
base, is obtained by choosing the initial vector so that the reference state
$(s_1,\ldots,s_m)=(0,\ldots,0)$ carries weight one, and all other states
carry weight zero. The whole state space for the given sector is constructed
automatically in the iterative process. For the ground state sector, the total
number of states is found to be given by Eq.~(\ref{Hcoefs}); this constitutes
a useful check of the numerical algorithm. Excited sectors, corresponding to
non-horizontal bases, can be similarly accessed by choosing another
appropriate reference state as the initial vector.

The above construction has enabled us to numerically diagonalize $T_m$ for $m
\le 27$. The corresponding eigenvalues are tabulated in Table~\ref{tab1}.

\subsection{Second construction}

The second construction of the transfer matrix is described here for the
original trimer tiling problem. We first define the tiling at level $n$ as
the set of all tiles that have at least one square with $y$-coordinate
less than or equal to $n$. And $H_n(x)$ is the height profile for this set
of tiles above $y=n$. Then $H_n(x)$ clearly lies between $0$ and $2$,
and may be characterized by a sequence of the type $21222000122110\cdots$.
The transfer matrix $T_{C,C'}$ is 1, if the height configuration $C'$ can
be reached from $C$ by adding some tiles, otherwise $0$.

The transfer matrix is constructed simply as follows: Let the configuration
$C$ be given by some sequence of $0$'s, $1$'s and $2$'s as above.  We add trimers to
this configuration to all sites with height $0$. We can place a
horizontal trimer at any place where three consecutive sites have height
$0$. This changes the heights at these sites to $1$. We place zero
or more horizontal trimers this way. To each remaining site with height $0$
we add a vertical trimer, so that the height of that site becomes $3$.
Now there are no sites with height $0$, and the maximum height is $3$. 
Finally we measure heights from a new reference point one unit higher, 
and decrease all heights by $1$. This gives the new height configuration 
$C'$ with heights given again by a sequence of $0$'s, $1$'s and $2$'s.

There is a simple way to represent this transfer matrix as a spin 
hamiltonian. Denote the three possible heights at site $i$ by a quantum 
spin that can be in any one of three orthonormal states $|0\rangle$, 
$|1\rangle$ and $|2\rangle$.  We denote the three states at site $i$ by 
$|0_i\rangle$, $|1_i\rangle$ and $|2_i\rangle$. Define the operator
$S^{-}_i$ by

\begin{equation}
S^{-}_i |h_i\rangle = | h'_i\rangle, \mbox{ where }  h-h' =1 \mbox{ (mod 3)}.
\end{equation}
And we define $P^{0}_i$ as the projection operator for the state 
$|0_i\rangle$, {\it i.e.}
\begin{equation}
P^{0}_i |0_i\rangle =|0_i\rangle, ~~P^{0}_i |1_i\rangle = 
P^{0}_{i}|2_i\rangle = 0.
\end{equation}
 
Then it is easy to see that the transfer matrix can be written as 

\begin{equation}
{\cal T}=  {\rm Tr}\prod_{i=1}^{L} 
\left[ 
\begin{array}{ccc}
S^{-}_i  & P^{0}_i & 0 \\
0 & 0 & P^{0}_i \\
P^{0}_i & 0 & 0 
\end{array}
\right]
\end{equation}

To prove this, we only need note that expanding the product, the only 
nonzero terms are of the form $S^- S^- P^0 P^0 P^0 S^- \ldots $, 
where we have a string  of $S^-$ at consecutive sites, 
interspersed with the product of three $P^0$'s at consecutive sites. 

As an example, let us consider random horizontal and vertical trimer
tilings of a square lattice with periodic boundary
conditions in the horizontal direction, \ie, an infinite cylinder of width $m$.
For convenience we choose $m$ to be a multiple of 3.
A typical tiling of the cylinder is shown in Fig.~\ref{figure1}.
It is easy to see that on a $3 \times \infty$ cylinder a horizontal  
trimer can be followed by three vertical  trimers, or  by 
another horizontal trimer in exactly three ways due to the
periodic conditions in the horizontal  direction.
The possible height
configurations are thus `000', `111', and `222', and
the $m=3$  transfer matrix is given by
\begin{eqnarray}
T = \left[ 
\begin{array}{cccccc}
3 & 1 & 0  \\
0 & 0 & 1  \\
1 & 0 & 0 
\end{array}
 \right] \,.
\end{eqnarray}

We can reduce the size of the transfer matrix by working in a sector where
the basis vectors are invariant under translations and reflections.
Thus for $6 \times \infty$ cylinder, in the sector where the irreducible 
word is empty,  we have six basis vectors,   
`000000', `000111', `111111', `000222', `111222', `222222', and the other 
vectors are related to these by symmetry. The number of vectors needed for
the transfer matrix for a $3n \times \infty$ cylinder is less than $H_{3n}$ 
by approximately a factor of $6 n$. The resulting size of the transfer matrix 
for  $n \le 8$  is shown in Table \ref{tab1}.
By using the rotational and translational symmetries the
reduction in the size of the transfer matrix (still in the ground state sector)
can be judged by comparing the 2nd and 3rd columns of Table \ref{tab1}.
The above construction has enabled us to numerically diagonalize $T_m$ for $m
\le 24$. 
It is less efficient than the sparse matrix obtained by
first construction but calculating the nonleading
eigenvalues is much easier this way.
The corresponding correlation lengths obtained from
the second eigenvalue are given in Table~\ref{tab1}.

We have also studied the case when $m$ is {\em not} a multiple of $3$.  
We recall that in the more familiar  case of {\em dimer} tilings, for
even $m$, there is a one-dimensional height mapping, and accordingly the
continuum limit is that of a free boson with $c=1$. For odd $m$, however,
the height representation has non-periodic boundary conditions, 
corresponding to a twist operator, which leads to renormalization of 
the effective central charge to  $c_{\rm eff} = -2$ \cite{Izm05}. 
Returning to the trimer problem, we see that taking $m \mbox{ (mod 3)} 
\neq 0$ also introduces  twist operators here, and would change the 
effective value of central charge. 
The entropy per site $S_m$ for $m \mbox{ (mod 3)} = 1$
and 2 is listed in Table \ref{tab1a}.

\begin{center}
\begin{table}[!h]
\begin{tabular}{|r|r|r|r|r|}
\hline
$m$ & ${\rm dim}_1(T_m)$ & ${\rm dim}_2(T_m)$ &  $S_m$ &  $\xi_2$ \\
\hline
3  & 3 & 3 &  0.37754275 & 0.58860147 \\ 
\hline
6  & 21 & 6 &  0.21764117 & 0.95122814 \\ 
\hline
9  & 183 & 19 &  0.18163298 & 1.50340426 \\ 
\hline
12  & 1\,773 & 99 &  0.17027036 & 2.15126321 \\ 
\hline
15  & 18\,303 & 672 &  0.16557863 & 2.83712631 \\ 
\hline
18  & 197\,157 & 5\,667 & 0.16322214 & 3.54463208 \\ 
\hline
21 & 2\,189\,799 & 52\,689 &  0.16187256 &  \\ 
\hline
24 & 24\,891\,741 & 520\,407 &  0.16102733 &  \\ 
\hline
27 & 288\,132\,303 &     &   0.16046299  & \\
\hline
\end{tabular} 
\caption{The dimension of the transfer matrices $T_m$, ${\rm
dim}_1(T_m)$, without symmetrization, 
and ${\rm dim}_2(T_m)$, with symmetrization,
is shown for different $m$.
Also shown are the entropy per site $S_m$  
and the correlation length $\xi_2$ defined in Eq.~(\ref{cor}).} \label{tab1}
\end{table}
\end{center}

\begin{table}
\begin{tabular}{|r|r|r|r|}
\hline
$m$ &  $S_m$ & $m$ & $S_m$  \\
\hline
4 & 0.15682941    & 5 &  0.13541741 \\
\hline
7 & 0.15773925   &  8 &  0.15031598 \\
\hline
10 & 0.15789911   & 11 &  0.15438343 \\
\hline
13 & 0.15806225   & 14 &  0.15603971 \\
\hline
16 & 0.15817988   & 17 &  0.15687242 \\
\hline
19 & 0.15825951   & 20 &  0.15734824 \\
\hline
22 & 0.15831410   & 23 &  0.15764475 \\
\hline
\end{tabular} 
\caption{Entropy per site $S_m$ for $m ({\rm mod} 3) = 1$
and 2.
} \label{tab1a}
\end{table}

\section{Numerical diagonalization of the transfer matrix}
\label{sec:diag}

Two-dimensional isotropic statistical systems with short-range
interactions at the critical point may
exhibit invariance under {\it conformal transformations}
\cite{Car87}.
For a cylinder of width $m$ the
finite-size corrections to the entropy per site $S_m$ are then of the form \cite{cFSS}
\begin{eqnarray}
S_m = S_\infty  + \frac{\pi c}{6 m^2} + o(m^{-2}) \label{concr}
\end{eqnarray}
where $S_\infty$ is the entropy per site in the thermodynamic limit $m \to \infty$ (i.e., in the infinite plane)  and $c$ is the {\it central charge} (conformal
anomaly number) which determines the universality class of the problem \cite{Car87}.

\begin{widetext}
\begin{center}
\begin{table}[!h]
\begin{tabular}{|r|r|r|r|r|r|r|r|r|}
\hline
$(m,m+3)$ & $S_{\infty}$ & $c$  & $(m,m+3)$ & $S_{\infty}$ & $c$  &
$(m,m+3)$ & $S_{\infty}$ & $c$  \\
\hline
(3,6)  & 0.16434064 & 3.664674  & (4,7) & 0.16728470 & -0.319490 & 
(5,8) & 0.16121121 & -1.231563 \\
\hline
(6,9)  & 0.15282642 & 4.456333   & (7,10) & 0.15818038 & -0.041283 & 
(8,11) & 0.15986633 & -1.167350 \\
\hline
(9,12)  & 0.15566128 & 4.017785   & (10,13) & 0.15805270 & -0.029334 & 
(11,14) & 0.15895040 & -1.055394 \\
\hline
(12,15)  & 0.15723777 & 3.584218   & (13,16) & 0.15829869 & -0.076314 & 
(14,17) & 0.15871184 & -1.000265 \\
\hline
(15,18)  & 0.15786649 &  3.314046  & (16,19) & 0.15840838 & -0.111719 & 
(17,20) & 0.15862738 & -0.968649 \\
\hline
(18,21)  & 0.15813526 &  3.147735  & (19,22) & 0.15845364 & -0.133847 & 
(20,23) & 0.15858708 & -0.946406 \\
\hline
(21,24)  & 0.15826623 &  3.037424  & & & & & & \\
\hline
(24,27)  & 0.15833844 & 2.957992 & & & & & & \\
\hline
\end{tabular} 
\caption{Estimates of entropy $S_{\infty}$ and the central charge $c$
from fits  using successive pairs of cylinder widths $m$.
} \label{tab2}
\end{table}
\end{center}
\end{widetext}

It is important that estimates of $S_\infty$ 
obtained by extrapolating of data for different values of $m \mbox{ (mod 3)}$
have to be consistent with each other.
Estimates for $S_\infty$ and $c$ obtained by fitting pairs
$(S_{m},S_{m+3})$ to Eq.~(\ref{concr}) are shown in Table~\ref{tab2}.
Clearly, there is still some residual $m$-dependence. 
In general, in
conformal field theory, one can also have a non-universal $1/m^d$
correction to scaling term in Eq.~(\ref{concr}),
\begin{equation}
S_m = S_\infty  + \frac{\pi c}{6 m^2} + \frac{e}{m^d} \label{threeterm},
\end{equation}
 with $2 < d \leq 4$.  Here we adopt the following strategy:
we choose trial values of $d$ and $S_{\infty}$ and obtain $c$ and $e$
from Eq.~\ref{threeterm} by sequential 2-point fits.
 We find that for different $m$ values there is reasonable convergence
for $d = 2.78 \pm 0.10$ and  $S_{\infty}= 0.158520 \pm 0.000015$.
In table \ref{tab2a},  we have listed the values of $c$ and $e$ 
obtained by two-term 
sequential fits for $S_m$ with $m =0, 1$ and $2 \mbox{ (mod 3)}$, using the 
fitting form Eq.~(\ref{threeterm}).  
The estimate for central charge is  $c = 2.15 \pm 0.2$.
Our estimate for the effective charge 
is $ c_{\rm eff} = -0.28 \pm 0.02 $ for $m = 1 \mbox{ (mod 3)}$, 
and $ c_{\rm eff} = -0.79 \pm 0.02$ for $m =2 \mbox{ (mod 3)}$.

\begin{widetext}
\begin{center}
\begin{table}[!h]
\begin{tabular}{|r|r|r|r|r|r|r|r|r|}
\hline
$(m,m+3)$ & $c$  & $e$ & $(m,m+3)$ & $c$  & $e$ & $(m,m+3)$ & $c$  & $e$ \\
\hline
(3,6) & 4.48761407  & -0.8859 & (4,7) &  -0.11255023  &   0.0932 & 
(5,8) & -0.77423123  & -0.5984 \\
\hline
(6,9) & 2.24822274  &  3.8068 & (7,10) & -0.26174392   &  0.4455 &
(8,11) & -0.78830675  & -0.5615 \\
\hline
(9,12)& 1.85242965  &  4.9419 & (10,13) & -0.27727375   & 0.4938 &
(11,14) &  -0.79454721  &  -0.5406 \\
\hline
(12,15)& 1.98106106  &  4.4810 & (13,16) & -0.27266918  &  0.4762 &
(14,17) & -0.79175635  & -0.5519 \\
\hline
(15,18)& 2.09438199  &  3.9984 & (16,19) & -0.27314362  &  0.4784 &
(17,20) & -0.78906173  & -0.5645 \\
\hline
(18,21)& 2.14525639  &  3.7489 & (19,22) & -0.27959514  &  0.5114 &
(20,23) &  -0.78903295  &  -0.5647 \\
\hline
(21,24)& 2.15734045  &  3.6821 & & & & & & \\
\hline
(24,27)& 2.14936018  &  3.7310 & & & & & & \\
\hline
\end{tabular} 
\caption{Estimates of $c$ and $e$ from 2-point fits with $e/m^d$ correction
for $d=2.774$ and $S_{\infty} = 0.15852$ for $m \mbox{ (mod 3)} =$ 0, 1
and 2.
} \label{tab2a}
\end{table}
\end{center}
\end{widetext}

\begin{figure}
\includegraphics[scale=.35, angle=270]{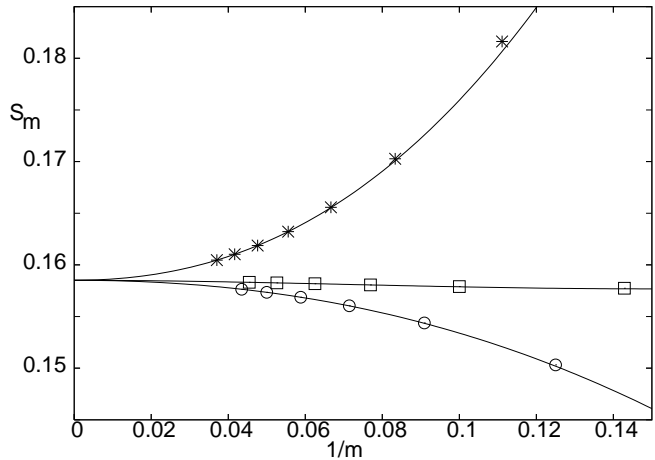} 
\caption{ Convergence of $S_m$ with $1/m$ for
$m ({\rm mod} 3) = 0$ ($\star$), 
$m ({\rm mod} 3) = 1$ ($\Box$), $ m ({\rm mod} 3)= 2$ ($\circ$).
Continuous lines are fits from Eq.~\ref{threeterm} that
converges to $S_{\infty} = 0.15852$ for
$d=2.774$ and $c$ and $e$ as listed in Table~\ref{tab2a}.
} \label{figure8}
\end{figure}

To inquire further into the critical behavior of trimer tilings, we can
measure as well $S^{(i)}_m = \frac{1}{m} \log \lambda_i(m)$, where
$\lambda_i(m)$ is some sub-leading eigenvalue ($i \neq 1$). 
From conformal field theory we expect to get corrections to follow the
behavior \cite{xFSS}
\begin{equation}
 S_\infty - S^{(i)}_m = \left( 2 \Delta_i - \frac{c}{6} \right)
 \frac{\pi}{m^2} + o(m^{-2}) \,,
 \label{xFSS}
\end{equation}
where $\Delta_i = h_i + \bar{h}_i$ is the scaling dimension of the field corresponding to the excited state described by $\lambda_i$.  Alternatively, Eq.~(\ref{xFSS}) may be stated in terms of the correlation length
\begin{eqnarray}
\xi_i = \frac{1}{\log{(\lambda_1/\lambda_i)}} \label{cor}
\end{eqnarray}
of the excitation on the cylinder. This then reads
\begin{eqnarray}
\xi_i = \frac{m}{2 \pi \Delta_i} \,, \label{concf}
\end{eqnarray}
i.e., the correlation length is proportional to the cylinder width.

We shall now give numerical evidence that Eq.~(\ref{xFSS}) holds true for
a number of excitations (i.e., that the resulting finite-size estimates
for $\Delta_i$ converge as $m \to \infty$). This is 
support for the hypothesis that trimer tilings are conformally invariant.
(We note that Kenyon has proved \cite{Ken01} that dimer tilings are
conformally invariant.)

\begin{figure} \includegraphics[scale=.8, angle=0]{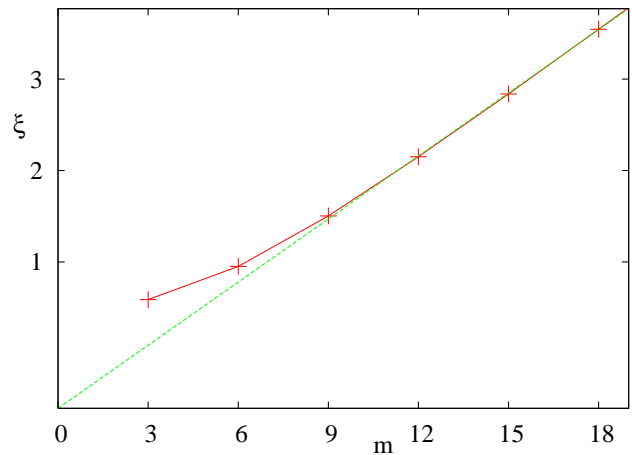} \caption{The
correlation length $\xi$ plotted against the cylinder width $m$. The
dotted line has the slope $a=0.23$. } \label{figure9} \end{figure}

The first kind of excitation $i=2$ in the ground state sector
(i.e., with the tilings grown on flat base) has already
been mentioned above. The corresponding correlation
length $\xi$ is shown in Table~\ref{tab1} and plotted against $m$ in
Fig.~\ref{figure9}. One finds a slope $a=\frac{1}{2\pi \Delta_2} \simeq
0.23$, corresponding to $\Delta_2 \sim 0.69$.

We have also studied the case of $3|m$, but with tilings grown from a
non-planar base. For simplicity we shall
consider only the two simplest sectors. The first one corresponds to
taking the initial state $s_1 = 1$, and $s_k=0$ for $k=2,\ldots,m$, in the
notation of Section~\ref{sec:1stTM}. The associated exponent $\Delta_{\rm
D}$ describes physically the decay of the correlation function between a
pair of widely separated dimer defects in the surrounding soup of trimers.
The second sector that we shall consider is built from the initial state
$s_1=1$, $s_2=2$, and $s_k=0$ for $k=3,\ldots,m$. The associated exponent
$\Delta_{\rm L}$ corresponds to a pair of L-shaped defects (or closely
bound compounds of a dimer and a monomer).

The estimates for $\Delta_{\rm D}$ and $\Delta_{\rm L}$, obtained by
adding a non-universal $1/m^d$ correction to Eq.~(\ref{xFSS}), are shown
in Table~\ref{tabdefects}. The dimer defect is the most relevant. In the
continuum limit, described by the two-dimensional height field, the
exponents $\Delta_{\rm D}$ and $\Delta_{\rm L}$ would describe the decay
of vortex-vortex correlation functions. Their final values appear to be of
the order $\Delta_{\rm D} = 0.62 \pm 0.1$ and 
$\Delta_{\rm L} = 0.82 \pm 0.1$, 
but clearly the words of caution on the slow convergence rate made
above when discussing the extrapolations of $c$ are also applicable here.

\begin{table}
\begin{tabular}{|r|r|r|r|r|}
\hline
$m$ & $S_m^{(D)}$ & $\Delta_{\rm D}$ & $S_m^{(L)}$ & $\Delta_{\rm L}$  \\
\hline
6 & 0.11284785 & & 0.10663883 & \\
\hline
9 & 0.13675337 & 0.511103 & 0.13215832 & 0.634525 \\
\hline
12 & 0.14560151 & 0.537242 & 0.14225002 & 0.684513 \\
\hline
15 & 0.14988060 & 0.559121 & 0.14733680 & 0.725974 \\
\hline
18 & 0.15229423 & 0.577165  & 0.15029069 & 0.761116 \\
\hline
21 & 0.15379778 & 0.592680  & 0.15217226 & 0.791808 \\
\hline
\end{tabular} 
\caption{Entropy per site for 
tiling of non-planar bases and estimates of scaling dimensions
with $1/m^d$ correction with $d=2.774$ correction.
} \label{tabdefects}
\end{table}

\section{Correlation functions}
\label{sec:corr}

Given a trimer tiling, we assign to each lattice face a state $s=1,2,\ldots,6$ according to its position
in the tiling. Our convention is that horizontal trimers are labeled as
 ~\fbox{1~~2~~3}~ and vertical trimers as
 ~\fbox{\parbox{0.2cm}{ 4\\ 5\\6}}~.

Then we can define the correlation function $G_{ij}({\bf x},{\bf Y})$
as the probability that the face at ${\bf X}$ is in the state $s=i$ and the
face at ${\bf Y}$ is in the state $s=j$ (with $1 \le i,j \le 6$).
In the thermodynamic limit, translational invariance implies that 
$G_{ij}({\bf X},{\bf Y})$ is only a function of $({\bf X}-{\bf Y})$, and 
we write it as $G_{ij}({\bf R})$ with ${\bf R}={\bf X}-{\bf Y}$.  
However, all the correlation
functions are not independent. For instance, we can express the function 
$G_{1 j}({\bf R})$ in terms of the function $G_{2 j}({\bf R})$ as
\begin{eqnarray}
G_{1j} ({\bf R}) = G_{2j}({\bf R}+ {\bf e}_x).
\end{eqnarray}
Using such equations, one can express the functions where $i$ or $j$ 
take values $1,3,4 $ or $6$ in terms of the functions 
$G_{2 2}, G_{2 5}$ and $G_{5 5}$. 
Again $G_{2 2}$ or $G_{5 5}$ are related by the symmetry in
the horizontal and vertical direction.
This only leaves the functions $G_{22}$ and $G_{25}$.
The conditional probability that the site ${\bf R}$ 
is occupied horizontally, given that
the origin is in state 2, is written $1/2 + f({\bf R})$,
so that the probability of finding a vertical trimer at ${\bf R}$
is $1/2 - f({\bf R})$. We therefore have
\begin{eqnarray}
G_{22} ({\bf R}) + G_{22}({\bf R} +{\bf e}_x) +  G_{22}({\bf R}-{\bf e}_x)
= \frac{1}{2} + f({\bf R}) \label{eqg22} \\
G_{25} ({\bf R}) + G_{25}({\bf R}+{\bf e}_y) +  G_{25}({\bf R}-{\bf e}_y)
= \frac{1}{2} - f({\bf R}) \label{eqg25}
\end{eqnarray}
The above equations can be solved as linear inhomogeneous equation in 
$G_{22}$ and $G_{25}$, and can be solved if $f({\bf R})$ is known
for all ${\bf R}$.
Thus we have only one unknown function $f({\bf R})$.
Note that Eq.~(\ref{eqg22}) is a difference equation 
in the $x$-coordinate (the $y$-coordinate does not change).
The general solution reads
\begin{eqnarray}
G_{22}(X,Y) = G_{22}^0(X) + A_Y \sin k_0 X + B_Y \cos k_0 X
\end{eqnarray}
where $k_0 = 2 \pi/3$.
But $A$ and $B$ must be zero, since the function $G_{22}(X,Y)$
is known to decay to a constant value $1/6$ for large $X$.
Thus we can determine $G_{22}$ and similarly $G_{25}$ uniquely
in terms of $f({\bf R})$.

\section{Monte-Carlo simulations}
\label{sec:mc}

We have already introduced the local move of a block of three vertical
trimers replacing a block of three horizontal trimers, or vice versa (see Fig.~\ref{figure4}).  It
has been shown that on a square lattice the above move is ergodic, \ie,
all possible configurations can be reached from any initial configuration
\cite{Ken92}. Starting from any initial configuration, say the standard
configuration with all trimers vertical, repeating the local operation at
randomly chosen sites generates a random trimer tiling.

In our  Monte Carlo simulation, we start with a 
$L \times L$ lattice fully packed with 
all trimers vertical ($L = 15, 45, 60, 90$). 
In one Monte Carlo step, we randomly select one of the trimers, and 
check if can be flipped using the move of Fig.~\ref{figure4}.  
If yes, the flip is made, if no, the 
move is rejected, and another site is selected. 
We  discard  the initial $10^5$ steps. 
Once the steady state is reached, we calculate the 
different correlation functions in the steady state. 
We generated data for  over  $10^5$ Monte-Carlo steps per site (MCS).

To verify that our sampling produces unbiased results, we have used
it to compute the average height on each of the nine different sublattices.
These can also be computed analytically as follows.
First, we define the absolute values of the heights by setting the mean height $\langle {\bf h} \rangle_1 = 0$
on sublattice 1. To compute $\langle {\bf h} \rangle_2$ on sublattice 2, we first note that
the height difference between site A, belonging to sublattice 1,
and site B, belonging to sublattice 2,
(i.e., along the horizontal edge joining 1 and 2 in Fig.~\ref{figure5}(a))
can either be 1 if the edge overlaps with the boundary of a
horizontal or vertical trimer, or it can be $-2$ or $2 \omega - \omega^2$
if it is one of the internal edges of a vertical trimer.
Hence the mean value of ${\bf h}_B - {\bf h}_A$
is $(1/6)(4 + 2 \omega - \omega^2 -2) = - \omega^2/2$, and since $B$ was arbitrary
$\langle {\bf h} \rangle_2 = -\omega^2/2$. 
In a similar way, we find that
\begin{eqnarray}
\langle {\bf h} \rangle_1 &=& \langle {\bf h} \rangle_5 = \langle {\bf h}
\rangle_9 = 0 \nonumber\\
\langle {\bf h} \rangle_2 &=& \langle {\bf h} \rangle_6 = \langle {\bf h}
\rangle_7 = - \omega^2/2 \nonumber\\
\langle {\bf h} \rangle_3 &=& \langle {\bf h} \rangle_4 = \langle {\bf h}
\rangle_8 =  \omega/2 \nonumber.
\end{eqnarray}
We have found these values of $\langle {\bf h} \rangle_2$ and $\langle {\bf h} \rangle_3$ to be
in excellent agreement with the numerical results.

\begin{figure}
\includegraphics[scale=.38, angle=270]{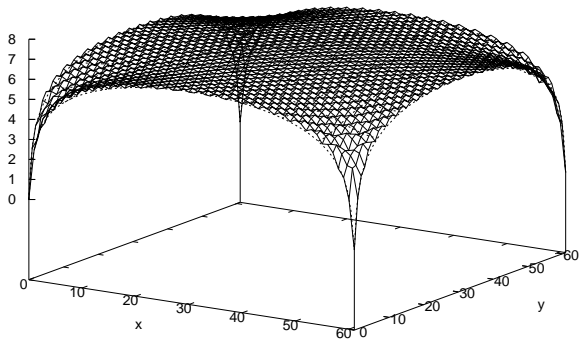} 
\caption{Correlation function H(x,y) for L=60.} \label{figure10}
\end{figure}

To compute the height-height correlation function 
\begin{eqnarray}
H(x,y) = \left\langle \left|{\bf h}(X+x,Y+y) - {\bf h}(X,Y)\right|^2 \right\rangle
\end{eqnarray}
we averaged the data over all positions $(X,Y)$. The results are shown
in Fig.~\ref{figure10}.
For next nearest neighbors, height difference squared
is 1 with probability 2/3,
4 and 7 each with probability 1/6.
Hence $H(1,0) = H(0,1) = 5/2$ which has been verified against the results
of the simulations.
For system size $L$, $H(x,y)$ varies for large $r$ as
\begin{eqnarray}
H(x,y) 
\approx \Delta_2 \log \left[ (B L)^2 \left\lbrace \sin^2 \left(\frac{\pi x}{L} \right) 
+ \sin^2 \left( \frac{\pi y}{L} \right) \right\rbrace + 1 \right]
\nonumber \\ 
\end{eqnarray}
where $\Delta_2 = 0.69$ and $B = 4$.
Since $H(1,1) \neq H(1,-1)$, there should also be a term depending
on $\sin^2( \pi x/L) -\sin^2( \pi y/L)$, but this should decay faster 
for large $r$,
and we have not included it in our fits.
For large $r = \sqrt{x^2 + y^2}$ it is easy to see that
\begin{eqnarray}
H(r) \approx 2 \Delta_2 \log (r)
\end{eqnarray}
or equivalently we have
\begin{eqnarray}
c - C(r) \approx  \Delta_2 \log (r)
\end{eqnarray}
where $c$ is a constant and 
$C(r) = \langle h(r) h(0) \rangle $.
From our estimates of $\Delta_2 = 0.69$ and $a = 0.23$, 
from Eq.~(\ref{concf}) it is seen
that $\Delta_2 = 1/ (2 \pi a)$ in agreement with the theory of
conformal invariance.

\begin{figure}
\includegraphics[scale=.35, angle=270]{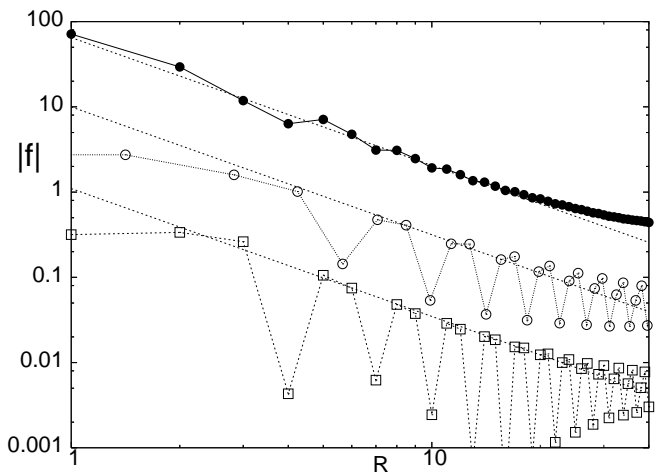} 
\caption{ $|f(R,\theta)|$ 
as functions of $R$ for $\theta = 0$ ($\cdot$), 
$\theta = \pi/4$ ($\circ$), $\theta = \pi/2$ ($\Box$).
Straight lines have slope -1.5.
} \label{figure11}
\end{figure}

In Fig.~\ref{figure11}, we have plotted $f(R,\theta)$ for three different directions 
$\theta = 0, \pi/4, \pi/2$. We used a lattice of size $90 \times 90 $, and 
$10^5$  
MCS to get the data.  In each case, we see that in each direction, the 
function $f(R, \theta)$ decreases as a power, $f \sim R^{-x}$, with $x
\approx 1.5$. However it is observed that the effective exponent decreases with $R$
and it is difficult to make meaningful estimates
due to large fluctuations in the data.
 
\section{Conclusions}
\label{sec:conclusion}

We have studied the problem of tiling the plane with trimers, and seen that it differs
from the well-known dimer tiling problem in a number of ways. First, the
number of sectors in the transfer matrix grows exponentially rather than linearly with
system size. Second, the natural height mapping has a target space dimension which is
two rather than one. If the tiling problem is conformally invariant, one could therefore
reasonably expect it to have central charge $c=2$. To check this hypothesis, and the
inference on $c$, we have performed extensive numerical transfer matrix calculations,
on cylinders of widths $m \le 27$ lattice spacings. The finite-size corrections exhibit
unexpectedly large correction-to-scaling terms, making difficult the precise estimations
of the critical exponents. The results are however clearly in favor of conformal invariance,
and this is confirmed by the logarithmic form of the height-height correlations as obtained
by Monte-Carlo simulations. Our estimate $c = 2.15 \pm 0.2$ is in marginal agreement with
the expectation $c=2$, but we cannot definitely rule out a more complicated behavior.
We have also measured numerically a number of other exponents, in particular the leading
thermal scaling dimension and the scaling dimensions related to pairs of geometrical defects (dimers and L-shaped trimers). \\

{\bf Acknowledgments.} This research has been supported by the
Indo-French Centre for the Promotion of Advanced Research (CEFIPRA),
project number 3402-2.

\appendix

\section{Irreducible words for a general tiling problem}

The working of section~IV can be adapted to a more general tiling problem in
which there are two types of straight tiles: horizontal $p$-mers (i.e., of
size $p \times 1$ elementary squares) and vertical $q$-mers (of size $1 \times
q$). We suppose for the rest of this appendix that $p \ge 3$ and that $q \ge
2$.

The words describing the sector decomposition of the transfer matrix are then
made of $q$ different letters. An irreducible word is one in which there is no
substring of $p$ consecutive equal letters. For $1 \le k \le p-1$ we then
define $N_k(n)$ as the number of irreducible words of length $n$ in which the
last $k$ letters are all equal, but in which the last $k+1$ letters are not
all equal.

The initial conditions for the recursion relations are by convention
\begin{equation}
 N_k(p-1) = 1 \mbox{ for } 1 \le k \le p-1 \,.
\end{equation}
This definition of $N_k(n)$ does not yet take into account the multiplicity
due to the $q$-dependent number of ways that one may chose the last $p-1$
letters. Denoting this multiplicity $M_k(p-1)$ one finds
\begin{eqnarray}
 M_k(p-1) &=& q^{p-1-k} (q-1) \mbox{ for } 1 \le k \le p-2 \nonumber \\
 M_{p-1}(p-1) &=& q \,.
\end{eqnarray}
Indeed, one needs to choose the first $p-1-k$ letters arbitrarily and then
complete the word with a single letter, chosen different from the last one
chosen arbitrarily. Note that the sum of all multiplicities is
$\sum_{k=1}^{p-1} M_k(p-1) = q^{p-1}$ as expected.

For any $n \ge p-1$ one then has the recursion relations
\begin{eqnarray}
 N_1(n+1) &=& (q-1) \sum_{k=1}^{p-2} q^{1-k} N_k(n) + q^{3-p} N_{p-1}(n) \nonumber \\
 N_k(n+1) &=& q N_{k-1}(n) \mbox{ for } 2 \le k \le p-2 \nonumber \\
 N_{p-1}(n+1) &=& (q-1) N_{p-2}(n) \,. \label{recrelgen}
\end{eqnarray}
This generalizes Eq.~(\ref{recursrels}). Equivalently, the recursion relations
can be written in matrix form by defining the vector $\vec{N}(n)$ with
elements $N_k(n)$. One then has $\vec{N}(n+1) = {\bf T} \vec{N}(n)$, where
${\bf T} = \{ T_{ij} \}$ is a matrix with elements that can be read off from
(\ref{recrelgen}). Iterating this, one finds
\begin{equation}
 \vec{N}(n) = {\bf T}^{n+1-p} \vec{1} = {\bf S} {\bf D}^{n+1-p} {\bf S}^{-1} \vec{1} \,,
\end{equation}
where $\vec{1} = [1,1,\ldots,1]^{\rm t}$ and ${\bf S}$ is the matrix of
eigenvectors of ${\bf T}$ that turns ${\bf T}$ into diagonal form through
${\bf D} = {\bf S}^{-1} {\bf T} {\bf S}$. Writing this out for given values of
$p$ and $q$ will yield an explicit solution generalizing Eq.~(\ref{solrecrel}).

The total number $t_n$ of irreducible words of length $n$ is then obtained by taking into account the 
multiplicities:
\begin{equation}
 t_n = \sum_{k=1}^{p-1} M_k(p-1) N_k(n)
\end{equation}
This generalizes Eq.~(\ref{multipli}).

To obtain the total number of words that are irreducible to the empty word, we
first define $G_\ell$ as the number of irreducible unfactorizable words with
initial letter $\ell=1,\ldots,q$. Introducing also
\begin{equation}
 G_0 = \sum_{n=0}^\infty \left( \sum_{\ell=2}^q G_\ell \right)^n
\end{equation}
the generalization of Eq.~(\ref{Ga}) reads
\begin{equation}
 G_\ell = \ell (G_0 \ell)^{p-1} \,.
\end{equation}
The corresponding generating function $g(x)$ then satisfies
\begin{equation}
 g(x) \left[ 1 - (q-1) g(x) \right]^{p-1} = x^p \,.
\end{equation}

The generating function for all words that are reducible to the empty word is
then
\begin{equation}
 H = \sum_{n=0}^\infty \left( \sum_{\ell=1}^q G_\ell \right)^n
   = \frac{1}{1-\sum_{\ell=1}^q G_\ell}
\end{equation}
and reads explicitly
\begin{equation}
 H(x) = \frac{1}{1-q \, g(x)} \,.
\end{equation}


\begin{thebibliography}{xx}
\bibitem{Ken00} R. Kenyon, \anp{28}{759}{2000}. Cylindrical boundary 
conditions for the trimer problem correspond to periodic boundary 
conditions for the height field only if the width $m$ is divisible by $3$.
\end{thebibliography}
\end{document}